\documentclass[a4paper]{jpconf}
\usepackage{graphicx}
\begin{document}
\title{Have Cherenkov telescopes detected a new light boson?}

\author{Marco Roncadelli}

\address{INFN, Sezione di Pavia, via A. Bassi 6, I -- 27100 Pavia, Italy}

\ead{marco.roncadelli@pv.infn.it}

\author{Alessandro De Angelis}

\address{Dipartimento di Fisica, Universit\`a di Udine, Via delle Scienze 208, I -- 33100 Udine,\\ 
and INAF and INFN, Sezioni di Trieste, Italy}

\ead{deangelis.alessandro@gmail.com}

\author{Oriana Mansutti}

\address{Dipartimento di Fisica, Universit\`a di Udine, Via delle Scienze 208, I -- 33100 Udine,\\ 
and INFN, Sezione di Trieste, Italy}

\ead{oriana.mansutti@gmail.com}

\begin{abstract}
Recent observations by H.E.S.S. and MAGIC strongly suggest that the Universe is more transparent to very-high-energy gamma rays than previously 
thought. We show that this fact can be reconciled with standard blazar emission models provided that photon oscillations into a very light Axion-Like Particle occur in extragalactic magnetic fields. A quantitative estimate of this effect indeed explains the observed data and in particular the spectrum of blazar 3C279. 
\end{abstract}

\section{Introduction}
So far, Imaging Atmospheric Cherenkov Telescopes (IACTs) have detected about 30 very-high-energy (VHE) blazars above $100 \, {\rm GeV}$ over distances ranging from the parsec scale for Galactic objects up to the Gigaparsec scale for extragalactic ones. By now, the fartest blazar observed by IACTs is 3C279 at redshift $z = 0.536$ detected by MAGIC. Given that these sources extend over a wide range of distances, not only can their intrinsic properties be inferred but also photon propagation over cosmological distances can be probed. This is particularly intriguing because VHE photons from distant sources scatter off soft background photons, thereby disappearing into $e^+ e^-$ pairs. Since the cross section $\sigma (\gamma \gamma \to e^+ e^-)$ peaks where the VHE photon energy $E$ and the background photon energy $\epsilon$ are related by $\epsilon \simeq (500 \, {\rm GeV}/E) \, {\rm eV}$, the horizon of the observable VHE Universe rapidly shrinks above $100 \, {\rm GeV}$ due to the presence of the Extragalactic Background Light (EBL) produced by galaxies during the whole cosmic history. In the presence of such an energy-dependent opacity, photon propagation is controlled by the photon mean free path ${\lambda}_{\gamma}(E)$ for $\gamma \gamma \to e^+ e^-$, and so the observed photon spectrum $\Phi_{\rm obs}(E,D)$ is related to the emitted one $\Phi_{\rm em}(E)$ by 
\begin{equation}
\label{a1}
\Phi_{\rm obs}(E,D) = e^{- D/{\lambda}_{\gamma}(E)} \ \Phi_{\rm em}(E)~.
\end{equation}
It turns out that ${\lambda}_{\gamma}(E)$ decreases like a power law from the Hubble radius $4.3 \, {\rm Gpc}$ around $100 \, {\rm GeV}$ to nearly $1 \, {\rm Mpc}$ around $100 \, {\rm TeV}$~\cite{CoppiAharonian}. Thus, Eq.~(\ref{a1}) implies that the observed flux is {\it exponentially} suppressed both at high energies and at large distances, so that sufficiently far-away sources become hardly visible in the VHE range and their observed spectrum should be {\it much steeper} than the emitted one.

Yet, the behaviour predicted by Eq.~(\ref{a1}) has {\it not} been detected by observations. A first indication in this direction was reported by the H.E.S.S. collaboration in connection with the discovery of the two blazars H2356-309 ($z = 0.165$) and 1ES1101-232 ($z = 0.186$) at $E \sim 1 \, {\rm TeV}$~\cite{aharonian:nature06}. Stronger evidence comes from the observation of blazar 3C279 ($z = 0.536$) at $E \sim 0.5 \, {\rm TeV}$ by the MAGIC collaboration~\cite{3c}. In particular, the signal from 3C279 collected by MAGIC in the region $E<220$ GeV has more or less the same statistical significance as the one in the range 220 GeV $< E <$ 600 GeV ($6.1 \sigma$ in the former case, $5.1 \sigma$ in the latter).

We argue that this circumstance strongly hints at the existence of a very light Axion-like particle (ALP), which is indeed predicted by many extensions of the Standard Model, including compactified Kaluza-Klein theories as well as in superstring theories~\cite{alp}. 

\section{DARMA scenario}

Our proposal -- to be referred to as the DARMA scenario -- can be sketched as follows~\cite{drm}. Because ALPs are characterized by a coupling to two photons, in the presence of an external magnetic field $\bf B$ the interaction eigenstates differ from the propagation eigenstates, so that photon-ALP oscillations show up (much in the same way as it happens for massive neutrinos). Photons are supposed to be emitted by a blazar in the usual way. In the presence of extragalactic magnetic fields -- whose existence is strongly suggested by AUGER observations~\cite{auger} -- some of them can turn into ALPs. Further, some of the produced ALPs can convert back into photons and ultimately be detected. In free space this would obviously produce a flux dimming. Remarkably enough, because of the EBL such a double conversion can make the observed flux {\it considerably larger} than in the standard situation! This is due to the fact that ALPs do {\it not} undergo EBL absorption. As a consequence, the observed photons travel a distance in excess of ${\lambda}_{\gamma}(E)$ and Eq. (\ref{a1}) becomes
\begin{equation}
\label{a1bis}
\Phi_{\rm obs}(E,D) = e^{- D/{\lambda}_{\gamma , {\rm eff}}(E)} \ \Phi_{\rm em}(E)~,
\end{equation}
which shows that even a {\it small} increase of ${\lambda}_{\gamma , {\rm eff}}(E)$ gives rise to a {\it large} enhancement of the observed flux. It turns out that the DARMA mechanism makes ${\lambda}_{\gamma , {\rm eff}}(E)$ shallower than ${\lambda}_{\gamma}(E)$ although it remains a decreasing function of 
$E$. So, the resulting observed spectrum is {\it much harder} than the one predicted by Eq. (\ref{a1}), thereby ensuring agreement with observations even for a {\it standard} emission spectrum. As a bonus, we get a natural explanation for the fact that only the most distant blazars would demand $\Phi_{\rm em}(E)$ to substantially depart from the emission spectrum predicted by the standard mechanism.

\section{Predicted energy spectrum}

Even though the existence of extragalactic magnetic fields ${\bf B}$ has been strongly suggested by AUGER observations~\cite{auger}, their morphology is largely unknown. So, it is usually supposed that they have a domain-like structure. It looks plausible to assume that $0.3  \, {\rm nG} < B < 1.0 \, {\rm nG}$ and that their coherence length $L_{\rm dom}$ is in the range $1\, {\rm Mpc} < L_{\rm dom} < 10 \, {\rm Mpc}$~\cite{dpr}.

We evaluate of the probability $P_{\gamma \to \gamma}(E,D)$ that a photon remains a photon after propagation from the source to us when allowance is made for photon-ALP oscillations as well as for photon absorption from the EBL, so that Eq. (\ref{a1bis}) gets replaced by
\begin{equation}
\label{a0as}
\Phi_{\rm obs}(E,D) = P_{\gamma \to \gamma}(E,D) \, \Phi_{\rm em}(E)~. 
\end{equation}

Our procedure is as follows. We first solve exactly the beam propagation equation over a single magnetic domain, assuming that the EBL is described by the ``best-fit model'' of Kneiske {\it et al.}~\cite{kneiske}. Starting with an unpolarized photon beam, we next propagate it by iterating the single-domain solution as many times as the number of domains crossed by the beam, taking each time a {\it random} value for the angle between ${\bf B}$ and a fixed overall fiducial direction. We repeat such a procedure $10^.000$ times and finally we average over all these realizations of the propagation process. 

As far as 3C279 is concerned, we find that about 13\% of the photons arrive to the Earth for $E = 500 \, {\rm GeV}$, representing an enhancement by a factor of about 20 with respect to the expected flux without DARMA mechanism (the comparison is made with the above ``best-fit model''). The same calculation gives a fraction of 76\% for $E = 100 \, {\rm GeV}$ (to be compared to 67\% without DARMA mechanism) and a fraction of 3.4\% for $E =  1 \, {\rm TeV}$ (to be compared to 0.0045\% without DARMA mechanism). The resulting spectrum is exhibited in Fig.~1. These conclusions hold for ALPs lighter than $10^{-10} \, {\rm eV}$ and for their two-photon coupling within the experimentally allowed range~\cite{drm}.

\begin{figure}[hb]
\centerline{\includegraphics[width=0.70\textwidth]{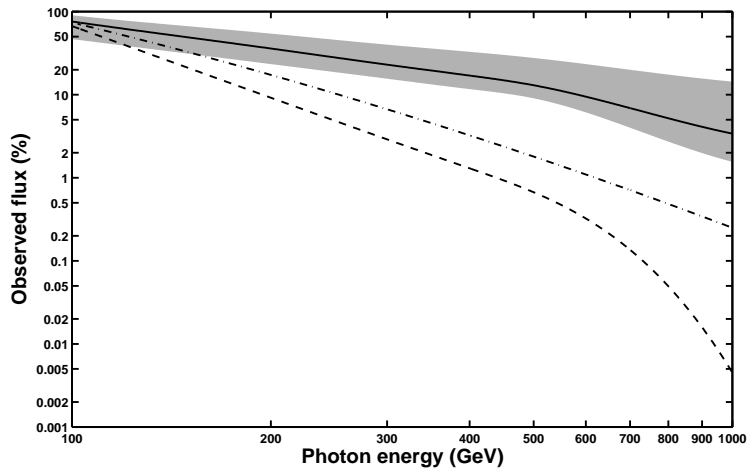}}
\caption{The two lowest lines give the fraction of photons surviving from 3C279 without the DARMA mechanism within the ``best-fit model'' of EBL (dashed line) and for the minimum EBL density compatible with cosmology (dashed-dotted line). The solid line represents the prediction of the DARMA scenario for  \mbox{$B \simeq 1 \, {\rm nG}$} and \mbox{$L_{\rm dom} \simeq 1 \, {\rm Mpc}$} and the gray band is the envelope of the results obtained by independently varying ${\bf B}$ and $L_{\rm dom}$ within a factor of 10 about such values.} \label{Fig:MV}
\end{figure}

\end{document}